\documentclass[11pt]{article}
\usepackage{graphicx,amsmath,color}
 \newcommand{\vtrail}{v_{\text{trail}}} 
 \newcommand{\vmax}{v_{\text{max}}}
 \newcommand{\vlead}{v_{\text{lead}}}
 \newcommand{\dif}[2]{\frac{\text{d}#1}{\text{d}#2}}
 \newcommand{\x}{$\times$}

\author{David Craft, Dualta McQuaid, Jeremiah Wala, Wei Chen, Ehsan Salari, Thomas Bortfeld}
\date{December 5, 2011}
\title{Multicriteria VMAT optimization}
\begin{document}
\maketitle
\thispagestyle{empty}

\begin{abstract} 

\noindent Purpose: 
To make the planning of volumetric modulated arc therapy (VMAT) faster and to 
explore the tradeoffs between planning objectives and delivery efficiency.

\noindent Methods:
A convex multicriteria dose optimization problem is solved for an angular grid of 180 equi-spaced
beams. This allows the planner to navigate the ideal dose distribution Pareto surface and select a plan
of desired target coverage versus organ at risk sparing.
The selected plan is then made VMAT deliverable
by a fluence map merging and sequencing algorithm, which combines neighboring 
fluence maps based on a similarity score and then delivers the merged maps together, 
simplifying delivery. Successive merges are made as long as the dose distribution quality 
is maintained. The complete algorithm is called {\sc vmerge}.

\noindent Results: 
{\sc vmerge} is applied to three cases: a prostate, a pancreas, and a brain. In each case, the 
selected Pareto-optimal plan is matched almost exactly with the VMAT merging routine, resulting in a high quality
plan delivered with a single arc in less than five minutes on average.

\noindent Conclusions:
{\sc vmerge} offers significant improvements over existing VMAT algorithms.
The first is the multicriteria planning aspect, which greatly speeds up 
planning time and allows the user to select the plan which represents the most desirable compromise 
between target coverage and organ at risk sparing. The second is the user-chosen epsilon-optimality 
guarantee of the final VMAT plan. Finally, the user can explore the tradeoff between delivery time and 
plan quality, which is a fundamental aspect of VMAT that cannot be easily investigated with current 
commercial planning systems. 

\end{abstract}

\section{Introduction}

In the late 1990s, intensity-modulated radiation therapy (IMRT) entered the clinical scene 
and quickly rose to a dominant position in radiation treatment.  
The first IMRT treatment was delivered as slice-by-slice rotational arc therapy using a special-purpose 
collimator \cite{carol95}. Soon after that, IMRT delivered from a small number (around 10) of fixed 
gantry angle positions using a general-purpose multileaf collimator (MLC) took over as the dominant 
IMRT delivery technique, and has been in that role ever since. Throughout this paper we will use the term ``IMRT'' to denote 
this specific delivery mode. The reader should keep in mind, however, 
that IMRT is a broader term that includes various types of delivery, 
including arc therapies. 


Rotational arc therapy in the form of {\em volumetric} intensity-modulated arc therapy (IMAT), delivered 
using an MLC, was proposed in 1995  \cite{yu1995}, but has been dormant for a long time until it recently 
re-emerged as volumetric modulated arc therapy (VMAT) with additional degrees of freedom, i.e., gantry speed and dose rate. One reason why it took so long for IMAT or VMAT to become a viable treatment option is that 
VMAT planning poses a larger optimization problem than IMRT because it delivers 
radiation from every angle around the patient, and therefore dose computations need to be done for 
far more angles than IMRT. An even bigger hurdle presented by VMAT optimization is due to the  coupling between adjacent angles. More specifically, to ensure an efficient VMAT delivery, one should not move the MLC leaves
more than needed between neighboring angles.
If VMAT plans are to be optimized with delivery time in mind, leaf positions need to be taken into account, which results in a large-scale non-convex optimization problem. Therefore, VMAT planning remains a great challenge and can be much more time consuming than IMRT planning \cite{oliver, rao}. 

We have previously introduced interactive multi-criteria-optimization (MCO) to facilitate 
IMRT planning \cite{kuefer-orspectrum}. 
The basic idea of interactive MCO is that a representative set of ``Pareto-optimal'' treatment plans 
are pre-calculated and stored in a database. Pareto-optimal treatment plans are those for which improving one criterion value is not possible unless some other criterion value deteriorates. The user can then navigate interactively through this 
database by exploring convex combinations of the database plans to obtain one that yields the desired tradeoff between different criteria. Therefore, MCO eliminates the time-consuming trial-and-error aspect of the planning process, and offers a more streamlined and effective approach to treatment plan optimization \cite{Craft2011}. Since VMAT poses a more challenging treatment planning problem, MCO has the potential for great benefit in VMAT planning. This has been our main motivation in developing this work. The goal of this paper is to develop and test a multi-criteria approach to VMAT planning that allows for exploring tradeoffs between planning objectives, as well as the dose quality and the delivery efficiency. We will achieve this by developing a two-stage method. In the first stage, MCO is used to navigate to a desired plan. In the second stage, the obtained plan is made VMAT deliverable, where the delivery efficiency is iteratively improved while maintaining the dose quality.

It is possible to blindly write down
the VMAT optimization problem and then apply various optimization algorithms to try to solve it. However,
due to the complexity of the problem, we feel it is more useful
to first clearly understand the physical basis (hardware, treatment dose parameters, etc.)
of the VMAT problem. 
The issues of fraction size, dose rate, leaf speed, and gantry speed are intertwined in VMAT.
Larger fraction sizes mean more dose delivered per degree, which means, all else being equal,
the gantry will rotate more slowly. This then implies that the leaves have relatively more time to move across the
field to create modulated fields. But at the outset, based on a certain fraction size and a set of machine 
parameters, should we expect to see generally small fields or large fields?  Can we be sure that when the gantry 
slows down or the dose rate decreases during a VMAT delivery that this is beneficial behavior
and not simply an artifact of the optimization? To address these questions, it is useful to consider
a specific scenario.

Assuming a 2 Gy fraction to the entire target can be delivered via a single 10 cm\x10 cm field,
this can be done in 200 monitor units (this is a typical MU/Gy calibration).
With a maximum dose rate of 600 MU per minute, the 2 Gy could be delivered in 20 seconds. 
At a maximum gantry speed of 6 degrees per second, 
a single revolution takes 60 seconds. Therefore, for a single revolution VMAT plan at maximal dose rate 
and gantry speed, in order to preserve the integral dose 
one will see average segment sizes on the rough order of $20/60 =1/3$
the size of a 10 cm\x10 cm field\footnote{In clinically implemented VMAT treatment 
planning systems, small segments are avoided by 
delivering treatment with a variable dose rate, 
which is typically much less than the maximum.}.
If one is willing to have the beam slow down to an average of half speed, completing the single arc 
in 120 seconds, one would see on average segments with a further 50\% reduction in size.
This provides some intuition on when to expect small segments in a VMAT delivery.

With a maximum leaf speed of 2.5 cm/sec, leaves can travel across a 10 cm field
in 4 seconds, in which time the gantry can rotate up to 24 degrees. To deliver highly modulated
fields that are spaced close together, the gantry may need to slow down. In general, in a VMAT delivery
the gantry needs to slow down when it takes longer to deliver the fluence pattern (required over 
some arc portion) than can be done at top gantry speed. It is useful to break this 
into two cases that represent the two
causes for gantry slowing. The first case is when the fluence map has fluence levels that exceed
the maximal fluence level that can be delivered at top gantry speed over the given arc portion.   
This clearly requires the gantry to slow down.
The second case is when the field is modulated so much that it takes more time than available
at top gantry speed to deliver all the isolated humps of the fluence map. 
The reason to distinguish these two cases is the following: in the first case, assuming the fluence 
profile is flat but at a large value, leaf speed is not the limiting factor because the gantry needs 
to slow down just to get enough dose in. For the second case, however, the modulated fields might be 
able to be delivered without slowing down the gantry if the leaves could move fast enough.

A useful relationship here is that the
delivery time for a fluence map to be delivered by a left-to-right leaf sweep across the field is equal
to the time it takes for the leaves to cross the field at top leaf speed plus the sum-of-positive-gradients
(SPG) term (see Equation \ref{seq-leafTime}). 
The SPG for an IMRT field is a measure of the ``ups and downs'' of the field. SPG can be minimized exactly in 
a convex optimization framework \cite{craft-spg}, whereas leaf travel distance, if incorporated
up-front in the optimization, results in a non-convex problem. In our approach, we handle leaf
travel issues in a VMAT-customized fluence map merging-and-sequencing routine which explicitly ensures that
the dose distribution quality is maintained, while the delivery efficiency is successively improved.
{\em Our algorithm is designed to solve one of the key design issues of VMAT planning: 
where to optimally slow down the gantry.} 
By merging like neighboring fluence maps and validating that the dose distribution
after the merge is still good, we eliminate unnecessary gantry slow downs which arise
from ``over-delivery'' of fluence maps. 
With our approach, the leaves travel back and forth at a
high frequency only when needed and likewise the beam slows down only when necessitated by leaf travel
requirements or SPG requirements.

VMAT treatments are currently delivered with Elekta \cite{Bedford2008} and Varian \cite{otto} 
equipment, and VMAT-like deliveries have been recently reported using Siemens equipment \cite{Salter2011}. 
The treatment delivery systems deployed by the different manufacturers  
have different designs, and thus impose different delivery constraints in treatment planning. The Elekta 
and Varian linacs both allow dynamic machine parameter changes during the irradiation, whereas for 
Siemens, the delivery proceeds via a burst mode in a step and shoot fashion. A full review of VMAT optimization techniques is provided by Yu~\cite{yu-vmat-review}. 
Here we review two algorithms which show the two general approaches used for VMAT planning.
In 2007, Varian adopted a one step algorithm for single-arc VMAT, reported by Otto~\cite{otto}, 
under the tradename RapidArc (Varian Medical Systems, Palo Alto, CA). The method first optimizes the MLC motions for a 
coarse sampling of static points. Finer sampling is achieved by iteratively adding samples 
interpolated between existing static points until the desired sampling frequency is reached.
Throughout, leaf positions are modified by local random search. By permuting the machine parameters 
directly, the Otto et al. solution to VMAT must solve a highly nonconvex problem. An alternative algorithm 
described in Bzdusek et al. \cite{Bzdusek}, that became commercialized under the name SmartArc (Philips 
Radiation Oncology Systems, Fitchburg, WI), uses a two-step approach to the problem. The algorithm initially
optimizes a set of fluence maps at a coarse angular sampling around an arc, followed by the sequencing of 
these maps and a direct parameter optimization of the resulting leaf positions. The initial fluence 
optimization step is a convex problem (if convex objective functions are used) and so can benefit from fast reliable 
optimization methods. Similar two-step approaches are found in \cite{amrt, klink}.
 
The method proposed herein uses a two-step approach similar to SmartArc: a fluence based optimization 
followed by leaf sequencing. However, our algorithm begins with fine beam spacing and 
leaf sequencing is accomplished using a unidirectional sequencing algorithm. This initial finely sampled plan
represents the best possible treatment given unrestricted time to deliver it. After obtaining this 
initial plan, neighboring fluence maps are iteratively merged to increase gantry speed and decrease delivery 
time. In this way, we work from the ideal solution towards one that is epsilon close (user-chosen) to dose 
optimality, but has greatly increased delivery efficiency. We call the complete algorithm {\sc vmerge}.

\section{Methods}

We begin by solving a 180 equi-spaced beam IMRT problem. 
We solve a multicriteria version of the IMRT optimization problem, which 
allows the planner to explore the tradeoffs between target coverage and healthy organ sparing, finally
choosing a best-compromise solution \cite{Craft2011, monz, tickletoes}.
Such a solution represents an ideal dosimetric plan, where treatment time is ignored.
To actually deliver this solution, one would deliver the full IMRT fluence maps at 
every 2 degrees, which would be time consuming. Instead, 
we successively coarsen this 180 beam fluence map solution
such that the delivery is made faster while the dose quality is 
kept within user selected bounds. Thus, in the sequencing step, we allow the user to explore the 
tradeoff between dose quality and delivery time.

In the following sections, we describe the details of each of these components of our VMAT 
planning approach.

\subsection{180 beam IMRT solution and Pareto surface plan selection} 

We consider the following multicriteria IMRT problem:

\begin{eqnarray}
\hbox{optimize~~~} & \{g_1(d),~g_2(d), \ldots g_N(d)\} \nonumber\\
\hbox{subject~to~~~}   & d = Df \nonumber\\
~~~~ & d \in C \nonumber\\
~ &  f \ge 0
\label{mco}
\end{eqnarray}
\noindent Here $d$ is the vector of voxel doses, $D$ is the dose-influence matrix, and $f$ is 
a concatenation of all the fluence maps into a single beamlet fluence vector. The constraint set $C$
is a convex set of dose constraints. This can include, for example, bounds on mean structure doses,
and minimum and maximum doses to individual voxels. 
Note, the constraints in this set should be known to be simultaneously
achievable (i.e they are {\em easy} constraints to meet). More difficult dosimetric goals, and mutually conflicting
goals, should be cast as objectives, in order to eliminate the possibility of infeasibility in this stage 1 optimization problem. 

The objective functions are $g_1(d),\ldots,g_N(d)$ where $N$ is the number of objectives defined.
The optimization objectives can be any of the following: minimize the 
maximum structure dose, maximize the minimum structure dose, or minimize or maximize the
mean structure dose. In general, any convex functions would be permissible \cite{romeijn}. For our optimization, we only consider these ones since they can be handled with a linear solver, and since in the multicriteria
planning context, they are typically
sufficient to create high quality treatment plans \cite{craft-alpha, hong}.

We solve this problem multiple times, approximating the Pareto surface, by following the methods
detailed in \cite{wei-proj}. Briefly, this method uses a feasibility projection solver that
iteratively projects onto violated constraints until all constraints are satisfied. Objectives
are turned into constraints with initially loose bounds which are gradually tightened until
they are within user specified tolerance of optimality. After the projection solver
runs for the $N$ objectives and some mixed objective plans (for mean dose objectives,
a mixed plan would be obtained by optimizing a weighted combination of the mean structure doses; 
for more sophisticated strategies, see \cite{wei-proj, bokrantz, rennen, craft-pgen}),
the user navigates the solution space, which amounts
to choosing the most preferable convex combination of the calculated Pareto surface plans. 
The best choice plan is user dependent. Indeed, this is the entire point of MCO: if optimal
plan selection criteria could be determined {\em a priori} then navigating a Pareto surface
would be unnecessary. For the cases studied herein, we selected plans with conformal dose distributions
to the targets and which emphasized large mean dose sparing of the most important critical structures.
The selected plan, which we consider the ideal dosimetric plan, is then passed to the leaf sequencing
and merging routine, described in Sections \ref{leafseq} and \ref{merging}.

\subsubsection{Fluence map smoothing}
\label{methods_smoothing}
In the {\sc vmerge} algorithm, fluence maps will be combined based on similarity. 
Smoothing the fluence maps during optimization will reduce the noise level in the fluence maps, therefore
making the similarity comparisons more meaningful. Fluence map smoothing also results in 
faster final delivery time, as the results will show. We consider two smoothing methods. The first
constrains all fluence levels to be below a certain value. We call this max beamlet smoothing, and it 
is attractive because it is simple to implement.
The second smoothing method uses an SPG smoother during the solver's projection steps. During the projection  
iterations, an SPG smoothing step is periodically called. For each fluence map, this step identifies the single row 
with the largest SPG and redistributes the fluences by reducing the peak fluences of that row by a factor of 0.99,
and then adding the 1\% to the neighboring adjacent beamlets. This is a heuristic approach to controlling
the SPG with a projection solver, inspired by smoothing kernels in projection-based image 
reconstruction \cite{gaborbook}. 
Note, both max beamlet smoothing and SPG smoothing are convex in the sense
that averaging two or more plans that have been smoothed will result in a plan that is as good as or better than
(in terms of the smoothing metric) the average of the smoothing metric of the base plans.

\subsection{Unidirectional leaf sequencing}
\label{leafseq}
In leaf sequencing, the task is to create a set of MLC leaf trajectories which produce the desired fluence 
map while the gantry rotates over the arc portion allotted to that map. 
To be deliverable, the 
leaf trajectories must not have leaf velocities greater than a given maximum value, either within the 
delivery of a given fluence map or between the delivery of one map and the next. A simple way of ensuring that 
this condition is met is to sequence the trajectories as an alternating sequence of left to right and right
to left dynamic MLC (dMLC) leaf sweeps. All leaves are aligned at one edge of the field at the beginning of
the arc portion delivery and align at the opposite edge of the field at the end of the arc portion,
ready to commence the next arc portion with the leaves moving in the opposite direction. 
By aligning the leaves at the edge of the field at the beginning and end of the arc portion, the gap 
between closed leaves can be hidden underneath a collimator jaw, thus reducing the undesirable leakage 
radiation. However, the effects of minimum leaf gaps and leaf leakage are not explicitly considered in this work.   

The dMLC leaf sweep trajectory, also known as ``sliding window'', 
is calculated using the equations provided by \cite{svensson94, stein94, spirou94}, which give the leaf velocity of the leading 
($\vlead$) and trailing ($\vtrail$) leaves in terms of the maximum leaf velocity ($\vmax$) and 
the local fluence gradient $\dif{f(x)}{x}$. 
The equations (\ref{seq-leafVel}) give the leaf 
velocities and require a constant dose rate ($r$) over the arc portion. 

\begin{eqnarray}
\left( \vlead(x)=\vmax,~\vtrail(x)=\frac{\vmax}{1+\vmax\dif{f(x)}{x}\frac{1}{r}}\right) & \mathrm{if~~}\,\dif{f(x)}{x}\geq0 \nonumber \\
\left( \vtrail(x)=\vmax,~ \vlead(x)=\frac{\vmax}{1-\vmax\dif{f(x)}{x}\frac{1}{r}}\right) & \mathrm{otherwise}
\label{seq-leafVel}
\end{eqnarray}

The time for all leaf pairs to traverse the field is given by (\ref{seq-leafTime}) and is governed by the 
width of the field $W_{F}$ and the SPG of the fluence map (which is equal to the SPG of the leaf row with highest SPG) 
divided by the dose rate $r$:

\begin{equation}
T=\frac{W_{F}}{\vmax}+\frac{\mathrm{max}_\mathrm{rows}\left(\sum\dif{f(x)}{x}^{+}\right)}{r}
\label{seq-leafTime}
\end{equation}
The $(\cdot)^{+}$ operator is shorthand for max$(\cdot, 0)$. That is,  
the gradient term is included in the sum only if it is positive.

Each fluence map is locked to a given portion of the gantry rotation arc, so that if the gantry rotation time over 
the arc portion is less than the leaf travel time required, the gantry speed is reduced. 
In a similar manner, if the required leaf travel time is less than the gantry rotation time, then 
the leaf travel time is increased by reducing the dose rate over the arc portion. It should 
be noted that currently a continuously variable dose rate is assumed, but also that there is no strict requirement 
that the leaves take the full duration of the time available to complete the fluence modulation. 
However, it is dosimetrically favorable to reduce the dose rate if this can be 
accomplished without an effect on delivery time, as this will lead to larger 
beam apertures\footnote{Consider a uniform field. To deliver this using the sliding window
technique when the dose rate is halved, the trailing leaf 
will wait twice as long before departing, resulting in a larger average 
distance between the leading and trailing leaves. In general, the fact that smaller dose rates 
lead to larger average leaf opening sizes can be shown by using Equations \ref{seq-leafVel}
to compute the leaf trajectory curves, and noting that the average opening size 
is equal to the area between these curves divided by the 
optimal delivery time, given in Equation \ref{seq-leafTime}.}. 
All else being equal, larger apertures are preferred due to smaller output factor 
variations at larger field sizes, making leaf end calibration less critical.

The leaf velocities of the leading and trailing leaves are then assigned to the right and left leaves respectively. 
The next fluence map is processed in the opposite direction (right to left) and the leading and trailing leaf 
trajectories assigned to the left and right leaves respectively. This process then repeats for all the fluence maps
in the VMAT arc. 

\subsection{Merging neighboring fluence maps}
\label{merging}
The purpose of the merging algorithm is to lower the beam-on treatment time by reducing the number of 
distinct fluence maps that need to be delivered. To deliver each fluence map, the leaves must make a 
full unidirectional sweep across the aperture over the arc portion that the fluence map is specified for. 
VMAT solutions with a large number of distinct fluence maps thus require the gantry to move slowly 
in order to give the leaves sufficient time to move across the fields.
Our merging algorithm iteratively merges neighboring fluence maps, reducing the total leaf travel 
distance and allowing the gantry to move more quickly around the full arc.

We begin with 180 optimized fluence maps which are delivered over the ranges $[0,2^\circ],[2^\circ,4^\circ],\\
...[358^\circ, 360^\circ]$. The initial solution is a high-quality treatment plan, and we seek to merge fluence maps in a way that preserves this optimized dose distribution. Our merging strategy is based on the observations that 1) merging fluence maps with the greatest degree of similarity will have the least effect on the final dose distribution, and 2) merging fluence maps with smaller arc portions will have less of an effect on the dose distribution than merging fluence maps defined over longer arc portions. 

These two observations allow us to define a similarity metric between any two neighboring fluence maps $f^1$ and $f^2$, with arc portion lengths of $ \theta_1$ and $ \theta_2$. The similarity metric $\delta$ is defined as the Frobenius norm of the difference between the maps (normalized by their arc portion lengths to make them comparable), and scaled by the combined arc portion length $\theta_1$ + $\theta_2$:
\begin{eqnarray}
  \delta(f^1,f^2) &=&\left( \theta_1+ \theta_2\right) \sqrt{\displaystyle\sum_{i,j}\left(\frac{f^1_{ij}}{ \theta_1}-\frac{f^2_{ij}}{\theta_2}\right)^2}
\end{eqnarray}
\noindent Here, $i$ and $j$ are indices over the rows and columns of the fluence maps. 
We incorporate this similarity metric into a greedy algorithm that merges a single pair of fluence maps with 
every iteration, such that after $n$ iterations the number of fluence maps is $180-n$. At each step, the 
neighboring pair with the lowest $\delta$ score is selected for merging. The merged fluence map is defined 
as the sum of the two neighboring fluence maps, with a new arc portion equal to the union of the initial two 
arc portions. In this merging strategy, each merge effectively averages two fluence maps over their combined arc length. 

The merged maps are then sequenced using a left-to-right sweep (or right-to-left, depending on which side the 
leaves begin on). As the gantry rotates through the arc and the leaves sweep across the field, the fluence from 
the leaf opening is deposited onto the nearest original $2^\circ$ fluence map. In this way we 
accurately (within $2^\circ$) simulate the fluence as it is delivered by the moving gantry and the moving MLCs.

The stopping criterion for the greedy search will depend on the planner's 
desired balance between treatment time and plan quality. It is expected that this tradeoff should lean heavily in 
favor of plan quality, so that the final merged dose distribution should be nearly identical to the original 180 beam VMAT plan. 
The final plan will be more time efficient than the 180 beam solution, but retain optimality to within a 
small factor epsilon, which can be chosen by the user.

\subsection{VMAT settings}
We use the following VMAT delivery parameters: maximum gantry speed = 1 rotation/min, maximum leaf speed = 
2.5 cm/sec, maximum dose rate = 600 MU/min. For each case we design a 2 Gy fraction plan. It is 
important to note that unlike step and shoot IMRT optimization,
where fraction dose scaling does not fundamentally affect the plan, here fraction dose is important since it is 
linked to dose rate, gantry speed, and leaf speed. For display purposes,
we scale the dose-volume-histograms (DVHs) up to the total dose delivered from all the fractions. We also
display the optimization formulations for the total dose.

\subsection{Clinical Cases}
We apply  {\sc vmerge} to three different clinical cases: a 
prostate case (voxel size: 3\x3\x2.5 mm$^3$, 1 cm\x1 cm beamlets) planned to 77 Gy, 
a pancreas case (voxel size: 2.6\x2.6\x2.5 mm$^3$, 1 cm\x1 cm beamlets) to 50.4 Gy, 
and a brain case (voxel size: 1.35\x1.35\x1.25 mm$^3$, 0.5 cm\x0.5 cm beamlets) to 32 Gy. 

We use CERR 3.0 beta 3 \cite{CERR} quadratic infinite beam (QIB) 
dose computation to calculate the 180 beam dose influence matrix, 
which takes about 5 minutes for the prostate case, 10 minutes for the pancreas case, and 20 minutes for the  
brain case. With the dose matrix calculated, dose distribution computation for a given fluence map takes seconds. 
The 180 beam optimization runs take between 1 and 5 minutes, depending on the treatment plan complexity 
and the size of the dose influence matrix.

In prostate treatment, the main dosimetric tradeoff is between the rectum dose and the target coverage.
However, for our analysis we hold the prostate coverage fixed at the prescription level, 77 Gy,
and consider the tradeoff between mean dose to the rectum (more precisely, the anterior rectum as contoured
by the physician), the bladder, and the unclassified tissue.

Pancreas is a challenging
site for VMAT for the same reason that it is challenging for beam angle optimization: the target volume is surrounded by kidneys,
stomach, and liver, and the optimal radiation entry directions are not obvious \cite{woudstra}. 

We select the brain case due to the potential of VMAT to treat isolated metastic lesions in a single 
gantry revolution. Minimizing treatment time in these cases is important because of the
precise set up required for such treatments. In this case there are two adjacent lesions, 
one of them being inside the brainstem, and the prescription to both is 32 Gy. For the optimization, we consider the targets 
as a single combined structure.

\noindent {\bf Prostate formulation}
\begin{eqnarray}
\hbox{minimize~~~} & \{\hbox{mean~rectum~dose},~\hbox{mean~bladder~dose},~\hbox{mean~u.t.~dose}\} \nonumber\\
\hbox{subject~to~~~}   & d = Df \nonumber\\
~~~~ & d_i \ge 77 \hbox{~Gy},~ \forall i \in \hbox{~target}\nonumber\\
~~~~ & d_i \le 45 \hbox{~Gy},~ \forall i \in \hbox{~femoral~heads}\nonumber\\
~~~~ & d_i \le 77*1.10 \hbox{~Gy},~ \forall i \nonumber\\
~ &   f \ge 0
\label{mco-prostate}
\end{eqnarray}
\noindent where u.t. stands for unclassified tissue: all the voxels not belonging to any other structure.

\bigskip

\noindent {\bf Pancreas formulation}
\begin{eqnarray}
\hbox{minimize~~~} &\{ \hbox{mean~shell~dose},~\hbox{mean~kidneys~dose},~\hbox{mean~liver~dose}, \nonumber\\
 \hbox{~~~~~~~~~~~~} & ~\hbox{mean~stomach~dose},~\hbox{-min~target~dose} \}~~~~~~~~~~ \nonumber\\
\hbox{subject~to~~~}   & d = Df \nonumber\\
~~~~ & d_i \ge 50.4*0.95 \hbox{~Gy},~ \forall i \in \hbox{~target}\nonumber\\
~~~~ & d_i \le 45 \hbox{~Gy},~ \forall i \in \hbox{~spinal~cord}\nonumber\\
~~~~ & d_i \le 50.4*1.18 \hbox{~Gy},~ \forall i \nonumber\\
~ &   f \ge 0
\label{mco-pancreas}
\end{eqnarray}
\noindent The shell is a 0.7 cm band around the target used to promote dose conformality to the target.

\bigskip

\noindent {\bf Brain formulation}
\begin{eqnarray}
\hbox{minimize~~~} & \{\hbox{mean~chiasm~dose},~\hbox{mean~brainstem~dose},~\hbox{mean~u.t.~dose},~\hbox{-min~target~dose}\} \nonumber\\
\hbox{subject~to~~~}   & d = Df \nonumber\\
~~~~ & d_i \ge 32*0.95 \hbox{~Gy},~ \forall i \in \hbox{~target}\nonumber\\
~~~~ & d_i \le 32*1.18 \hbox{~Gy},~ \forall i \nonumber\\
~ &   f \ge 0
\label{mco-brain}
\end{eqnarray}

\subsection{Sensitivity to algorithm settings}
For the prostate case, we
investigate some variations to the algorithm. The first is smoothing, discussed in Section \ref{methods_smoothing}, 
where we compare no smoothing,
max beamlet smoothing, and SPG smoothing. We also investigate how the final VMAT solution 
depends on beamlet size and beam angle spacing, in order to show that our solution technique
yields a fundamentally correct VMAT plan and not one that is sensitive to parameter choices.
Since most commercial VMAT solutions calculate the final dose on a 2 degree gantry spacing, we choose
to use this as our baseline angular spacing grid. We examine the nature of the VMAT solution that arises
when this grid is coarsened to 4 degrees (thus, we start by solving a 90 beam IMRT problem).
We then take this angular grid and further investigate shrinking the beamlet size by a factor of 2 in the
leaf travel direction (creating 0.5 cm\x1 cm beamlets). 

\section{Results}

\subsection{Prostate case}

We navigate to a solution with (mean rectum dose, mean bladder dose, mean u.t.) = (0.50, 0.50, 0.12)*77 Gy. 
Plan quality is assessed using the mean dose to the anterior rectum, femoral heads, and bladder, 
the standard error from the prescription level of the dose to the prostate, 
and the volume of the prostate receiving the prescription dose.

\begin{figure}[ht] 
\centerline{\includegraphics[width=6.5in]{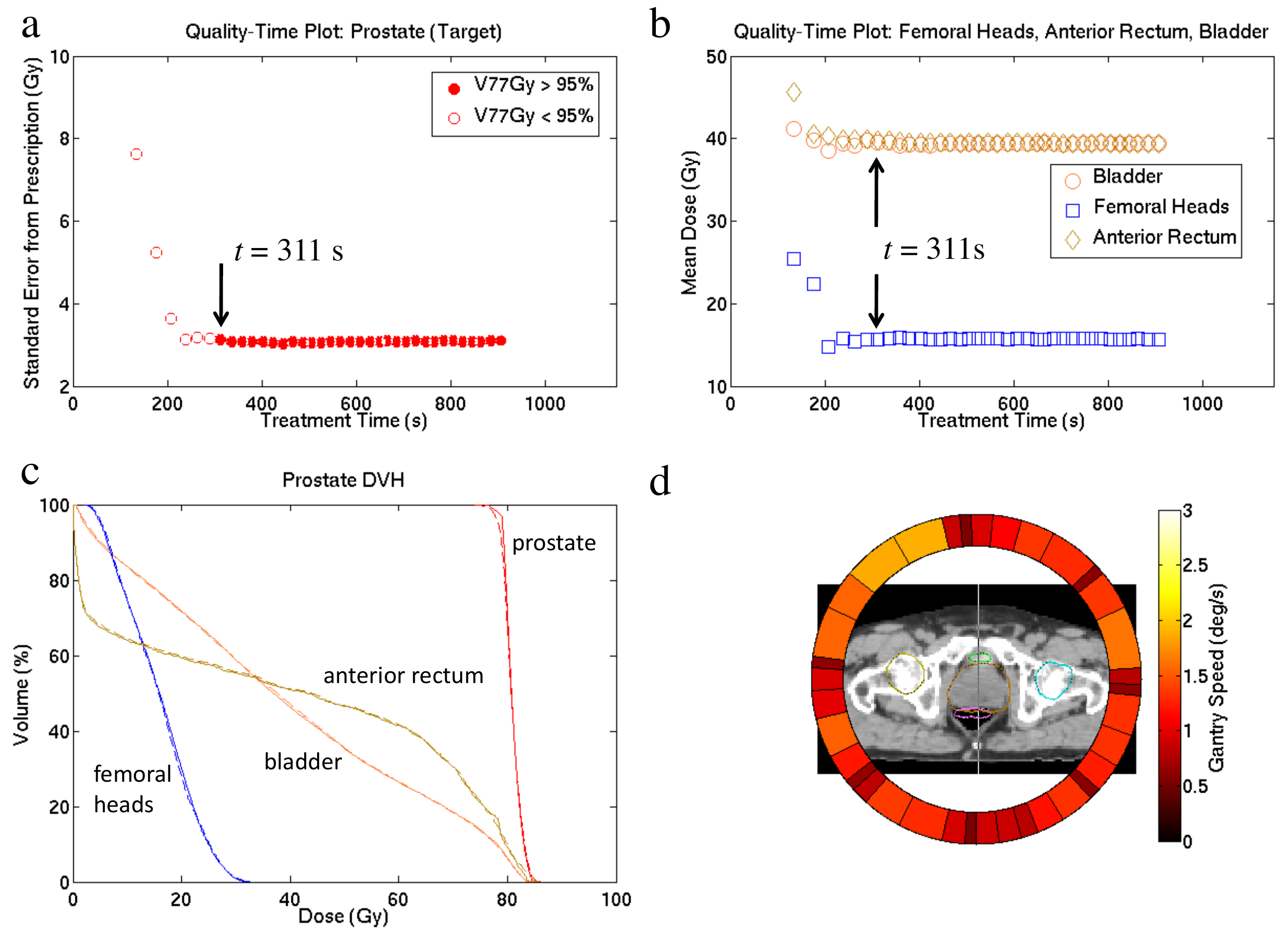}}
\caption{Results of the  {\sc vmerge} algorithm for prostate VMAT. The merged plan that was determined 
to have the best tradeoff between quality and treatment time is indicated by the black arrow. 
Quality-time tradeoff plots are shown for the a) prostate target and b) the bladder, femoral 
heads and anterior rectum. c) The DVH data for the original (solid) and merged (dashed) plan. d) The arc portion plot for the merged plan, showing the gantry speed at different angles.}
\label{wpprostate}
\end{figure}

Figure~\ref{wpprostate} shows the results of  {\sc vmerge} for the prostate case. 
The plan does not degrade until after 145 iterations, where we begin to lose target 
coverage. 
Therefore, we select a plan requiring 35 arc portions and a treatment time of 311 s 
(original time of 906 s with 180 arc portions) as having the optimal tradeoff between 
treatment time and plan quality.
This tradeoff between quality and
time is shown for both the target (Figure~\ref{wpprostate}a) and three key 
organs-at-risk (Figure~\ref{wpprostate}b). The DVH comparing the original plan (solid line) 
to the simplified plan (dashed line) is shown in Figure~\ref{wpprostate}c. The simplified 
plan has (mean rectum dose, mean bladder dose) = (0.50, 0.50)*77 Gy, with 95.3\% of the prostate 
volume receiving the full dose (V77Gy). Femoral head constraints are 
easily maintained. The plot of the arc portions in Figure~\ref{wpprostate}d shows that the gantry speed 
varies during the course of the single arc. Final arc portion sizes range from $4^\circ$ to $22^\circ$. 
Smaller arc portions require the gantry to slow down to allow the leaves time to traverse the aperture, and 
represent the fluence maps that are most dissimilar to their neighbors.

\subsubsection{Sensitivity: fluence map smoothing}
\begin{figure}[ht] 
\centerline{\includegraphics[width=6.5in]{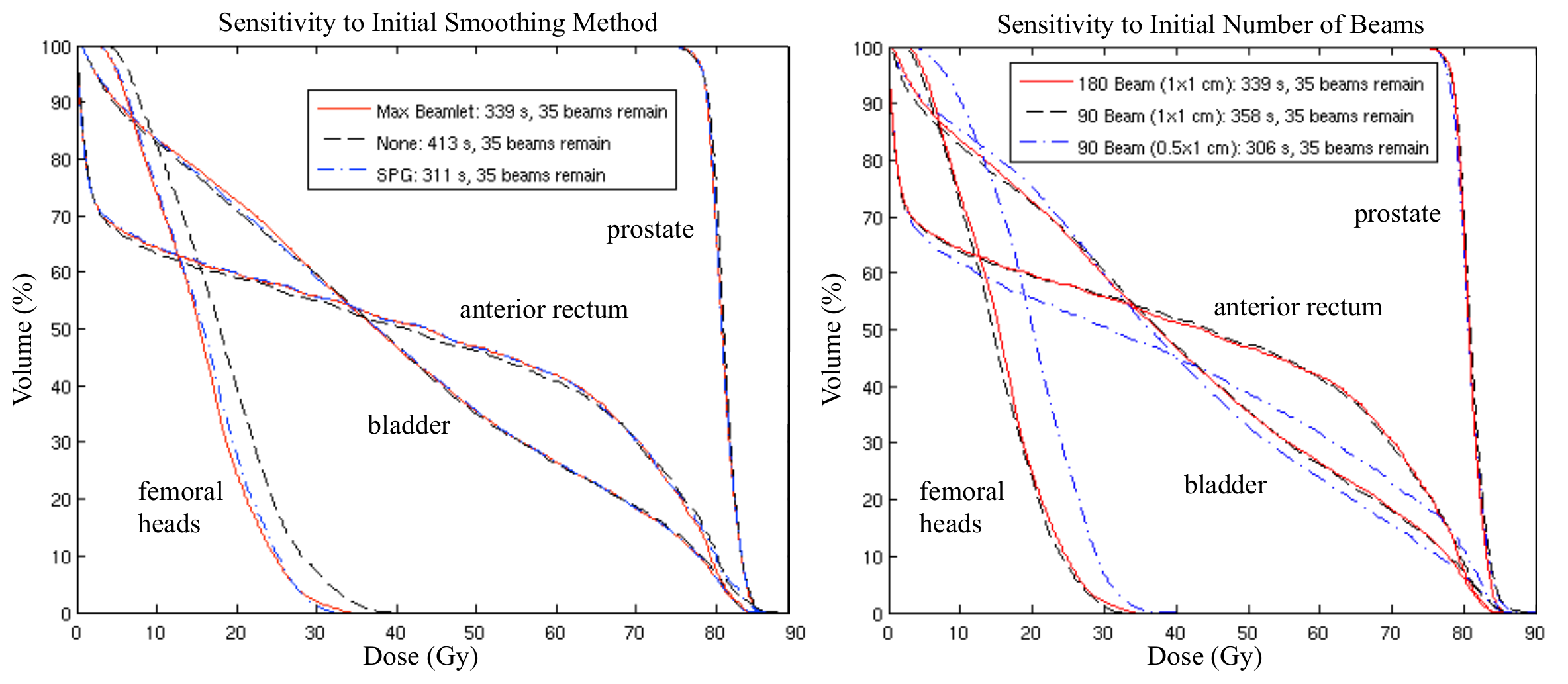}}
\caption{The sensitivity of {\sc vmerge} to different initial plans was tested on our prostate case. a) DVH plots 
for the final merged plans for 3 different 180 beam initial plans: max beamlet smoothing, no smoothing, and SPG 
smoothing.  b) DVH plots for the final merged plans for 3 different initial plans, all with max beamlet 
smoothing: 180 beams, 90 beams, and 90 beams with small beamlets (0.5 cm\x1 cm).}
\label{smooth}
\end{figure}

For the prostate case, the final treatment delivery time  
for no smoothing, max beamlet smoothing, and SPG smoothing, 
are 413, 339, and 311 seconds respectively. Figure~\ref{smooth}a shows the DVH plots for the final 
merged plans for these three smoothing methods. Each plan is created by merging from an initial 180 
beam solution, and represents the point on the tradeoff curve with the best compromise between 
plan quality and treatment time. The three plans are highly similar, differing only slightly in the
dose to the femoral heads, which is simply reflective of the original plan (since all plans easily satisfy
the femoral head constraint of 45 Gy, feasible plans exist with different femoral head doses;
in this case the initial 180 beam ``no smoothing'' plan happened to have a larger femoral head dose).  
Because SPG smoothing offers a significant improvement on the delivery time,
we use it for all of the nominal cases (Figures \ref{wpprostate}, \ref{wppancreas} and \ref{wpbrain}).

\subsubsection{Sensitivity: discretization settings}
Figure~\ref{smooth}b shows DVH plots for final plans created by iterative merging 
from an initial 180 beam solution (1 cm\x1 cm beamlet), 
90 beam solution (1 cm\x1 cm), and 90 beam solution with reduced beamlet sizes (0.5 cm\x1 cm). The final plans 
have similar treatment times and target coverage. In the optimization with the smaller beamlets, 
we reduce the rectum dose down as much as possible.
As expected, finer beamlet resolution yields better dose distribution shaping (the mean 
anterior rectum dose is reduced from 50\% to 44\% of the prescription dose), which indicates the 
value of reducing the beamlet size (for a theoretical analysis of the effect of beamlet size, see \cite{beamletsize}). 
But the stability and usefulness of the algorithm is unchanged
by beamlet size. To ease the computational burden we remain with the 1 cm$^2$ beamlets for the pancreas case. 
We switch to 0.5 cm$^2$ beamlets for the brain case due to its overall smaller target. 
This beamlet resolution is available in, for example, the Varian Millennium 120 MLC.

\subsection{Pancreas case}

For the pancreas case we navigate to a solution with (mean shell dose, 
mean kidneys dose, mean liver dose, mean stomach dose) = (0.93, 0.24, 0.22, 0.21)*50.4 Gy. 

\begin{figure}[ht] 
\centerline{\includegraphics[width=6.2in]{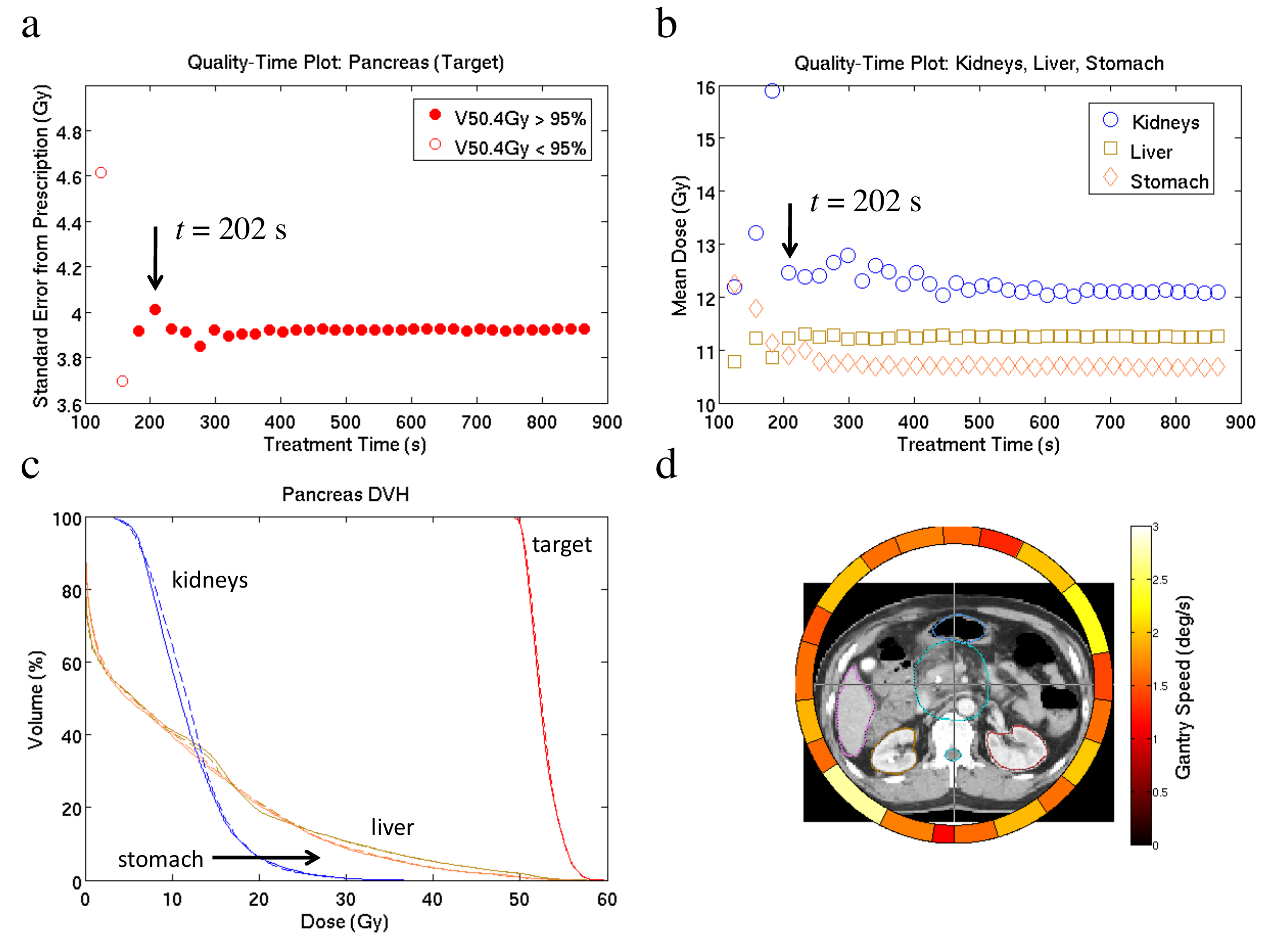}}
\caption{Results of the {\sc vmerge} algorithm for pancreas VMAT. The merged plan that was determined to have the best tradeoff between quality and treatment time is indicated by the black arrow. Quality-time tradeoff plots are shown for the a) pancreas target and b) kidneys, liver and stomach. c) The DVH data for the original (solid) and merged (dashed) plan. d) The arc portion plot for the merged plan, showing the gantry speed at different angles.}
\label{wppancreas}
\end{figure}

Figure~\ref{wppancreas} shows the {\sc vmerge} results. Plan quality begins to degrade after 160 merges, after which the mean dose to the kidneys, liver and stomach begins to increase. The selected plan requires 20 fluence maps and 202 s (initial time of 859 s), and reflects nearly the same DVH as the original 180 beam solution. 
Final arc portion sizes range from $8^\circ$ to $28^\circ$. 
The selected plan has (mean kidneys dose, mean liver dose, mean stomach dose) = (0.24, 0.22, 0.21)*50.4 Gy, with 98.3\% of the tumor volume receiving the full dose (V50.4Gy).

\begin{figure}[!ht] 
\centerline{\includegraphics[width=6.4in]{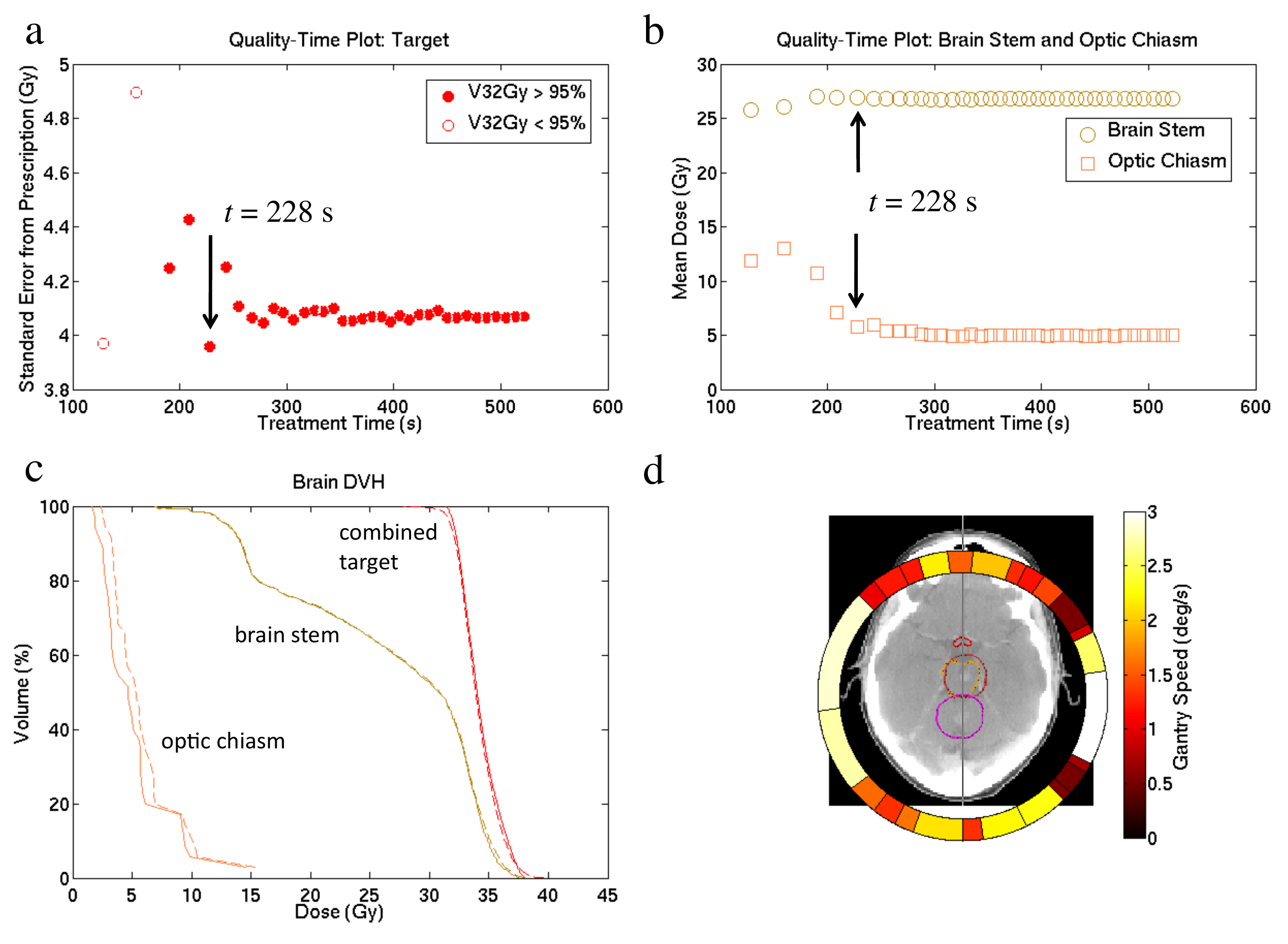}}
\caption{Results of the {\sc vmerge} algorithm for brain VMAT. The merged plan that was determined to have the best tradeoff between quality and treatment time is indicated by the black arrow. Quality-time tradeoff plots are shown for the a) tumor target and b) the brain stem and optic chiasm. c) The DVH data for the original (solid) and merged (dashed) plan. d) The arc portion plot for the merged plan, showing the gantry speed at different angles.}
\label{wpbrain}
\end{figure}

\subsection{Brain case}
For the brain case we navigate to a solution with (mean chiasm dose, mean brainstem dose, mean u.t. dose) = (0.15, 0.9, 0.44)*32 Gy and all target voxels receiving 32 Gy or higher.  

Figure~\ref{wpbrain} shows the {\sc vmerge} results. Plan quality begins to degrade after 155 merges, after which the mean dose to the optic chiasm begins to rise. The selected plan requires 25 fluence maps and 228 s (initial time of 522 s), and reflects nearly the same DVH as the original 180 beam solution. The selected plan has (mean chasm dose, mean brainstem dose) = (0.17, 0.90)*32 Gy, with 
97.6\% of the target volume receiving the full dose (V32Gy). 
Figure~\ref{wpbrain}d shows that from the anterior and posterior angles, finer sampling is required to 
shape the dose around the optic chiasm. The smallest arc portion sizes are $4^\circ$. 
Maximal gantry speeds are achieved near the lateral angles, where the radiation does not have to conform 
around an OAR, and the 
largest arc portion is $50^\circ$.
 
\section{Discussion and Conclusions}
{\sc vmerge} is designed to intrinsically address two closely related
VMAT issues: 1) how much fluence modulation is needed as the gantry proceeds around the patient, 
and 2) with what frequency do the fluence maps need to be delivered. The second point
is an important one in the context of minimizing delivery time in the finite leaf speed setting.
Finite leaf speed is the single parameter that makes the single arc coplanar VMAT optimization problem so challenging.
As leaf speed approaches infinity, the complexity and delivery time of a plan is governed solely
by the SPG of the fluence maps, and this quantity can be minimized exactly in a convex optimization 
setting \cite{craft-spg}. 
However, in the finite leaf speed setting, for fields even with low SPG the gantry may have 
to slow down just to give the leaves time to travel across the field.
This information cannot be represented in a convex optimization setting. 
Similar to the decomposition of IMRT planning into a convex
fluence map optimization step and then a leaf sequencing step, we show that
we can decompose the VMAT problem into two sequential stages, with the additional challenge arising 
from the continuous motion of the leaves and the gantry.

We adopt an approach of starting with a fine solution and gradually coarsening it. By successively coarsening,
we decrease the delivery time while maintaining a good dose distribution. Given current computing capacity 
and tailored algorithms to solve the IMRT problem (e.g. \cite{wei-proj,gpu-imrt}), 
it is not difficult to solve a 180 beam IMRT optimization problem. 
Our method of starting from the fine ideal solution and then making
it deliverable contrasts with all of the other VMAT approaches which start from a coarse solution
and then add segments to improve the dose quality \cite{otto, yu-vmat-review}. Another way to view 
the difference is: we start at the global minimum of a relaxed convex problem and then modify that 
solution to make it deliverable while staying as close as desired to the ideal plan, 
while the coarse-to-fine approaches first employ a global search and 
then use local refinements to optimize further, but do not yield any guarantees on 
optimality. Such guarantees are important both for high quality patient care and for proper comparisons
of treatment modalities, comparisons which can easily be muddled when using planning systems with no 
optimality guarantees.  

We study two approaches to create initially smoother 180 beam IMRT fluence maps.
Neither of the two methods has a dramatic effect on the final dose distribution, but both offer
improvements of the final treatment time.  Max beamlet smoothing reduces
treatment time from 413 seconds to 339 seconds (i.e. by 74 seconds, an 18\% reduction), and SPG smoothing 
by another 28 seconds. Even without smoothing, the merging routine combines neighboring fluence maps,
so noise in the maps that is not dosimetrically meaningful or useful will tend to get washed out. 
Still, having initially smooth maps allows the merging routine to merge down to a solution with improved delivery time.
For practical implementation, max beamlet smoothing is trivial to implement in the projection solver.
SPG smoothing on the other hand, while able to be written as a convex minimization problem,
is implemented heuristically as described above. An exact minimization of 
SPG can be done by introducing auxiliary variables \cite{craft-spg}, but this has not been pursued at this time. However, 
due to the usefulness of SPG smoothing, which is even greater for higher dose-per-fraction plans (where the SPG 
term in equation \ref{seq-leafTime} becomes more dominant), it is prudent to continue to investigate how 
to efficiently minimize SPG for large IMRT problem instances.

Using a finely spaced angular grid of beams and an IMRT solver to compute optimal fluence maps, one 
may observe that neighboring maps are highly similar. Indeed, in symmetric cases like a donut-shaped target
with a central circular organ at risk, the optimal maps from each angle are identical\footnote{In general, 
adjacent fluence maps of the 180 beam IMRT solution 
will be similar due to the underlying similarity of the dose-influence matrices, which can be explained by the 
small angle approximation. More precisely, for dose-influence similarity to lead to fluence map 
similarity one requires that the optimizer uses the dose-influence
matrix in a consistent way. Gradient algorithms, linear programming, and projection solvers are examples
of such methods.}. And yet, even with 180 identical fluence maps, one does not know from this information
how frequently those modulated fields need to be formed by the MLC leaves. {\sc vmerge} resolves this by 
successively merging similar neighboring maps for as long as the plan quality is maintained.
Even with our simple greedy selection strategy (merging the current two neighboring fluence maps 
with the best similarity score and employing no backtracking), we see that we can merge many of the maps 
and not change the plan perceptibly. A somewhat surprising result of our analyses, however, is the large number 
of adjacent merges that can be done
and still yield a good dose distribution. Examining Figures \ref{wpprostate}d and \ref{wpbrain}d, one can observe that
the largest angular sweeps are generally the sweeps which avoid the critical structures. 
In that case, any fluence modulation is due to target coverage rather than OAR sparing, so the relevant voxels
of the dose-influence matrix become the target voxels. Since target voxels are generally closer to iso-center than
OAR voxels, what can be considered a ``small angle'' is relatively larger for target voxels, 
which offers some justification for the large 
arc portions. For the pancreas case, Figure \ref{wppancreas}, the OARs
are spread evenly around the target, so the final arc portions are more uniform in size. It remains to be 
investigated if something other than the greedy merging strategy, and/or merging based on 
dose distribution similarities rather than fluence map similarities, could be more successful.

{\sc vmerge} is a two-stage method where the user first
explores the dosimetric tradeoffs exactly as done for IMRT MCO, 
and then either presses a button which does automatic merging to a pre-specified epsilon dose 
deviation threshold, or uses the merging routine to interactively explore the tradeoff between 
dose quality and delivery time, i.e. the information displayed in 
Figures \ref{wpprostate}, \ref{wppancreas}, and \ref{wpbrain}.
To create these figures, we run the merging algorithm until the gantry is moving at maximum
speed for the entire rotation, in order to show the entire plan quality-delivery time tradeoff curve.
This merging takes on the order of 5 minutes for each of the cases (this includes computing the dose and 
plan evaluation metrics after each merge; if you knew {\em a priori} how many merges you wanted, the merging 
would take on the order of 10 seconds). 
The planning time needed for a complete {\sc vmerge} planning session can be estimated as follows: dose-influence
matrix calculation (10 minutes) + computation of Pareto surface plans (15 minutes) + navigation (5 minutes) + 
merging and sequencing (5 minutes), which totals to 35 minutes. Dose-influence matrix calculation and
Pareto surface computation can each be parallelized, which would then speed the whole procedure to 
around 15 minutes.  While a direct comparison with commercial VMAT planning time is beyond the scope of this
report, the literature indicates times in the range of 20-60 minutes \cite{vplantime1, oliver}.

Although we used fairly simple formulations for the initial 180 beam IMRT solution, the method proposed herein 
does not preclude the use of more sophisticated formulations. For example, quadratic penalty formulations,
dose-volume constraints, equivalent uniform dose, and biological objective functions could all be used (provided
that memory and time efficient implementations exist for solving these other formulations).
Also, there is flexibility in the sequencing/merging step, and any number and type of user-defined cutoff values
could be defined to determine what solution along the delivery time/plan quality curve 
should be delivered. We have also used the simple pencil beam decomposition 
approach for the IMRT and VMAT problem. The dose distribution produced by such ``Dij'' approaches, in
na\"{\i}ve implementations, can degrade significantly when final dose calculations -- which include
output factor corrections and leaf transmissions -- are performed. However, we have proceeded with the 
Dij approach due to its mathematical tractability and because it is possible to successfully include 
delivery effects into Dij-based approaches \cite{jelen}.

Because we allow the gantry to slow down as much as possible to deliver the required fluence patterns,
there can be no dosimetric advantages to double arc (or more) solutions.
The only possible advantage for multiple arcs is treatment time.
Double arc solutions will be treatment time superior
when most of the fluence maps are highly modulated large fields which can be delivered
quicker in two sweeps, with the leaves reset on the second sweep to be in favorable
positions. For example, a double hump fluence map, with the humps separated by a wide zero
fluence section, would be faster with a two arc approach if the first arc delivered the first hump
and the leaves could be positioned correctly during the second arc to deliver the second hump.
To make a double arc plan overall faster in treatment delivery, one would need
a large number of the fluence maps to be of this nature.
We speculate that such situations will not arise often clinically, and that therefore single
arc solutions, with good optimizers, will typically be the right choice. On the other hand, 
a partial arc solution could save treatment
time by allowing the gantry to spend more time at angles that are more beneficial for radiation
delivery. A partial arc solution could be formed by prescribing an arc based on user experience, 
or by observing the original 180 beam solution and deciding to eliminate an angle sector which 
has generally low fluence. Once an arc is chosen, the multicriteria dose optimization and subsequent 
merging steps would proceed exactly as specified in this report. 


Other extensions to the proposed approach include collimator rotation, couch rotation,
and non unidirectional leaf sequencing. There could be a reduction in treatment time by aligning
the collimator to the fluence maps in such a way to minimize fluence map delivery time (Equation \ref{seq-leafTime}). 
The collimator angle trajectory could be determined by examining the fluence maps
of the 180 beam IMRT solution with this metric, and 
therefore seems not an overly difficult extension of the approach,
although one would need to implement a method for forming fluence maps with a rotating MLC collimator.
Couch rotation on the other hand, which leads to non-coplanar arcs, leads to a much larger 
optimization space and is accordingly a much more challenging problem. Finally, we have considered only 
unidirectional leaf sequencing in this paper. It is possible that more general 
leaf sequencing would allow one to deliver the merged maps more quickly. However, for fluence rows with 
multiple ups-and-downs, one always needs to sweep both leaves across almost the entire row to achieve 
inner row modulations. For this reason, we do not believe there will be dramatic gains in switching to more 
general leaf sequencing, but we do plan on investigating this.

While interesting algorithmic challenges remain for VMAT (non coplanar arcs, dynamic
collimator rotations, optimal partial arc creation, and integrating the MCO
dose exploration with the delivery efficiency tradeoff), we have introduced a method 
for single arc coplanar VMAT that guarantees delivery of an epsilon-optimal dose distribution.
Since {\sc vmerge} starts with 180 equi-spaced beams, and then makes it VMAT-deliverable, 
the method guarantees a provably optimal (up to an arbitrarily small user-specified tolerance)
treatment plan, which is something no commercial VMAT planning system currently does.

\vspace{.3cm}

\noindent {\bf Acknowledgements} The project described was supported by Award Number R01CA103904 from the National 
Cancer Institute. The content is solely the responsibility of the authors and 
does not necessarily represent the official views of the National Cancer Institute 
or the National Institutes of Health. 
Thanks to Brian Winey for providing the brain case.

\bigskip
\bibliographystyle{unsrt}
\bibliography{all}

\end{document}